\documentstyle[12pt]{article}
\newlength{\mathspace}
\def\t #1{\tilde{#1}}
\def\h #1{\hat{#1}}
\def\b #1{\bar{#1}}
\def\np#1{ Nucl. Phys. B#1}
\def\pr#1    { Phys. Rev. D#1 }
\def\pl#1{ Phys. Lett. B#1}

\def\ijmp#1  { Int. Jour. Mod. Phys. A#1 }
\def\mpl#1   { Mod. Phys. Lett. A#1 }
\def\begineq{\begin{equation}}
\def\endeq{\end{equation}}
\def\eqabegin{\begin{eqnarray}}
\def\eqaend{\end{eqnarray}}
\def\nn{\nonumber}

\begin{document}
\baselineskip=0.7cm
\setlength{\mathspace}{2.5mm}

%%%%%%%%%%%%%%%%%%%%%%%%%%%%%%%%%%%%%%%%%%%%%%%%%%%%%%%%%%

                                %titlepage

%%%%%%%%%%%%%%%%%%%%%%%%%%%%%%%%%%%%%%%%%%%%%%%%%%%%%%%%%%
\begin{titlepage}

   \begin{normalsize}
     \begin{flushright}
                 US-FT-35/96 \\
                 hep-th/9607157\\
     \end{flushright}
    \end{normalsize}
    \begin{LARGE}
       \vspace{1cm}
       \begin{center}
         {An Orbifold and an Orientifold of}\\ 
         {Type IIB Theory on K3 $\times$ K3 }\\ 
       \end{center}
    \end{LARGE}

  \vspace{5mm}

\begin{center}
           
             \vspace{.5cm}

            Shibaji R{\sc oy}
           \footnote{E-mail address:
              roy@gaes.usc.es}

                 \vspace{2mm}

        {\it Departamento de F\'\i sica de Part\'\i culas} \\
        {\it Universidade de Santiago,}\\
        {\it E-15706 Santiago de Compostela, Spain}\\
      \vspace{2.5cm}

    \begin{large} ABSTRACT \end{large}
        \par
\end{center}
 \begin{normalsize}
\ \ \ \
We consider the compactification of type IIB superstring theory on 
K3 $\times$ K3. We obtain the massless spectrum of the resulting two 
dimensional theory and show that the model is free of gravitational
anomaly. We then consider an orbifold and an orientifold projection
of the above model and find that their spectrum match identically and
are anomaly-free as well. This gives a dual pair of type IIB theory
in two dimensions and can be understood as a consequence of SL(2, Z)
symmetry of the ten dimensional theory. We also point out the M-theory
duals of the type IIB compactifications considered here. 
\end{normalsize}

\end{titlepage}
\vfil\eject

\begin{large}
\noindent{\bf I. Introduction:}
\end{large}
\vspace{.5cm}

Recent developments of string dualities [1--4] have enhanced our 
understanding of the
non-perturbative behavior of various string theories considerably. 
By assuming
the existence of an eleven dimensional quantum theory, known as M-theory
[3,5--11], the
string dualities can be understood more or less in a unified way with the 
exception of certain symmetries of type IIB theory in ten dimensions [12,13]. 
It has
been argued in ref.[13] that the symmetries of type IIB theory can be made
more transparent by postulating the existence of a twelve dimensional quantum
theory called as F-theory. Given the close relationships among F-theory, 
M-theory, and string theories, it has been pointed out in [15] that certain
four dimensional compactifications of F-theory are related to M-theory and
string theory compactifications below four dimensions. Thus, a deeper 
understanding of string theory compactifications in two dimensions will shed
light in our understanding of compactifications of F-theory in four dimensions
and holds the promise of solving the cosmological constant problem along the
lines proposed by Witten [16].

Having pointed out some motivations, we study in this paper a two dimensional
compactification of type IIB superstring theory on K3 $\times$ K3. Six 
dimensional compactifications of type IIB theory on K3 and its orientifold
have been studied before. In the first case [17], one gets a model with chiral 
N=2 supersymmetry (counting in terms of 6d Weyl spinors) which contains a
gravity multiplet alongwith twenty-one antisymmetric tensor multiplets as its
massless spectrum and is shown to be free of gravitational anomaly. This model
has been shown [18,19] to be equivalent to M-theory 
compactification on an orbifold
T$^5$/Z$_2$ where five tensor multiplets come from the untwisted sector and 
the remaining sixteen tensor multiplets come from the twisted sector by
placing sixteen Ramond-Ramond (R-R) five-branes on the internal space as 
dictated by the condition of gravitational anomaly 
cancellation. In the second case [20],
when one considers the orientifold projection of compactification of type IIB 
theory on K3, one gets a chiral N=1 supersymmetric model which contains in 
addition to a gravity multiplet, nine tensor multiplets and twelve 
hypermultiplets in the untwisted sector and eight vector as well as eight
hypermultiplets in the twisted sector by placing eight R-R five-branes on the
internal space. An identical spectrum can also be obtained [21,22] from 
an orbifold
model of M-theory on (K3 $\times$ S$^1$)/Z$_2$, but in this case eight tensor 
and eight hypermultiplets come from the twisted sector. Gravitational 
anomalies have been shown to be cancelled in all these models constructed. In 
fact,
in some cases the condition of gravitational anomaly cancellation 
provides a useful guide for
recovering the states in the twisted sector. In this paper, we extend the
previous cases and consider the compactification of type IIB theory further
on another K3. This two dimensional model will be shown to possess a chiral 
N=8 supersymmetry (counted in terms of 2d Majorana-Weyl (M-W) spinors). We
compute the complete massless spectrum of the model by making use of the
index theory and the Dolbeault cohomology of the simplest Calabi-Yau manifold
K3 [23]. We show that this gives a consistent 
compactification of type IIB theory 
in 2d, since the model is free of gravitational anomaly. We then project out
the above model by an orbifold group $\{1,\,(-1)^{F_L}\,\cdot\,\sigma\}$ and
an orientifold group $\{1,\,\Omega\,\cdot\,\sigma\}$. Here $F_L$ denotes the
space-time fermion number operator on the left moving sector of the 
world-sheet, $\sigma$ denotes the involution on the K3 surface and $\Omega$
is the orientation reversal of the world sheet. Type IIB theory is invariant 
under these operations. We compute the massless spectrum of this orbifold 
and the orientifold models both from the untwisted sector and from the 
twisted sector. We find that the spectrum, which has chiral N=4 supersymmetry,
match precisely for the two models
and thus we obtain a dual pair of type IIB theory in two dimensions. This can
be understood as a consequence of the SL(2, Z) symmetry of the ten dimensional
type IIB theory. M-theory duals of the type IIB compactification on
K3 $\times$ K3 as well as the orbifold and orientifold models, considered
here, have also been pointed out.

This paper is organized as follows. In section II, after briefly reviewing the
compactification of type IIB theory on K3, we compute the spectrum for the
K3 $\times$ K3 reduction. We also show that the model is anomaly-free. We 
describe the orbifold projection of this model in section III. We compute 
the spectrum of this orbifold model for both the untwisted and the twisted 
sectors and show that the model is anomaly-free as well. The orientifold 
projection is considered in section IV. In this case we find that the spectrum
matches identically with the previous orbifold model as a consequence of the
SL(2, Z) invariance of the ten dimensional theory. We also mention in brief
the M-theory duals of the two dimensional compactifications of type IIB theory
that are considered here. Finally, we present our conclusions in section V.

\vspace{1cm}

\begin{large}
\noindent{\bf II. Compactification of Type IIB Theory on K3 $\times$ K3:}
\end{large}

\vspace{.5cm}

The massless spectrum of type IIB theory in ten dimensions are given by the 
tensor product of a left-moving and a right-moving super Yang-Mills multiplet
of the same chirality as follows [24]:
\eqabegin
\bf(8_v + 8_c) \otimes (8_v + 8_c) &=& 
\bf(1 + 28 + 35_v + 1 + 28 + 35_-)_B\nn\\
& & \bf + (8_s + 8_s + 56_c + 56_c)_F
\eqaend
Here `{\bf v}', `{\bf s}' and `{\bf c}' represent the vector and two 
inequivalent spinor representations of SO(8). `{\bf B}' and `{\bf F}'
represent respectively the bosonic and the fermionic states. So, the bosonic
sector of this theory contains a graviton $(\bf 35_v)$, 
denoted by $\h g_{\h \mu 
\h \nu}(\h x)$\footnote[1]{We will denote the ten dimensional fields and
coordinates with a 
`hat', six-dimensional objects with a `tilde' and the two dimensional
fields without any accent. The objects on the internal manifold will be denoted
with a `bar'.} 
an antisymmetric tensor $(\bf 28)$, denoted by $\h B^{(1)}_
{\h \mu \h \nu}(\h x)$ and a dilaton $\bf(1)$, denoted by 
$\h \phi^{(1)}(\h x)$ from the Neveu-Schwarz--Neveu-Schwarz (NS-NS) sector
$\bf(8_v \otimes 8_v)$ and another scalar $\bf(1)$, denoted by 
$\h \phi^{(2)}(\h x)$, another antisymmetric tensor $\h B^{(2)}_
{\h \mu \h \nu}(\h x)$ $\bf(28)$ and a four-form antisymmetric tensor 
$\bf(35_-)$,
denoted by $\h A^-_{\h \mu \h \nu \h \rho \h \sigma}(\h x)$, whose 
field-strength is anti-self-dual, from the R-R sector $\bf(8_c \otimes 8_c)$.
In the fermionic sector, there are two spin 3/2 $\bf(56_c + 56_c)$ 
Rarita-Schwinger fields which are antichiral and are denoted as 
$\h \psi^{(1)\,-}_{\h \mu}(\h x)$ and $\h \psi^{(2)\,-}_{\h \mu}(\h x)$. Also,
there are two spin 1/2 $\bf(8_s + 8_s)$ chiral Majorana-Weyl fermions,
denoted as $\h \lambda^{(1)\,+}(\h x)$ and $\h \lambda^{(2)\,+}(\h x)$.
The two sets of spinors come from NS-R $\bf(8_v \otimes 8_c)$ and R-NS
$\bf(8_c \otimes 8_v)$ sectors.

We now briefly review the K3 reduction of type IIB theory to set up our
notations and conventions. A ten dimensional
field will be decomposed into a six dimensional (R$^6$) external field and
a four dimensional (K3) internal field as $\h \Phi (\h x) = \t {\Phi} (\t x)
\otimes \b \Phi (\b y)$ where $\t x^{\t \mu}$ is the coordinate on $R^6$ and 
$\b y^{\b m}$
is the coordinate on K3. $\t \Phi(\t x)$ will correspond to the massless
particle on $R^6$ if $\b \Phi(\b y)$ satisfies certain differential 
equations on the
internal manifold determined by the equations of motion of the ten dimensional
fields. Massless spectrum of the reduced theory then corresponds to the number 
of solutions of these differential equations. We also need to know certain 
properties of K3 surface, for example, it admits a Ricci-flat, anti-self-dual
Riemann curvature and has the following non-zero Betti numbers, $b_0 = b_4
= 1$, $b_2^+ = 3$, $b_2^- = 19$, where $b_2^+ (b_2^-)$ denotes the number of
self-dual (anti-self-dual) harmonic (1, 1) forms. Using this general strategy
and the properties of the K3 surface [25], 
we find that the ten dimensional metric  
$\h g_{\h \mu\h \nu}(\h x)$ yields one six dimensional graviton $\t g_{\t
\mu \t \nu}(\t x) \otimes \bf I$ (corresponding to one ($b_0 = 1$) constant
zero mode on K3), no gauge fields $\t A_{\t \mu}(\t x) \otimes \b f_{\b m}
(\b y)$
(since  $b_1 = 0$) and 58 scalars $\t a_1(\t x) \otimes \b h_{\b m \b n}
(\b y)$. Here 
$\b h_{\b m \b n}(\b y)$ satisfies the Lichnerowicz equation and the 
Lichnerowicz operator
has $b_2^+ \times b_2^- + 1 = 58$ zero modes [26,27] on 
K3 corresponding to the number
of gauge invariant deformations of the metric preserving the Ricci flatness 
condition. Similarly, from $\h B_{\h \mu \h \nu}^{(1)}(\h x)$ we get one  
antisymmetric tensor $\t B_{\t \mu \t \nu}^{(1)}(\t x) \otimes \bf I$ in six
dimensions and 22 scalars $\t b_1 (\t x) \otimes \b f_{\b m \b n}(\b y)$, 
since there are
22 harmonic two-forms on K3 altogether. The ten dimensional dilaton $\h \phi
^{(1)}(\h x)$ simply gives a scalar $\t \phi^{(1)}(\t x) \otimes \bf I$ in six
dimensions. Having thus described the reduction of the NS-NS sector, we now
turn to the R-R sector. The scalar $\h \phi
^{(2)}(\h x)$ gives a six dimensional scalar $\t \phi^{(2)}(\t x) 
\otimes \bf I$. The antisymmetric tensor $\h B_{\h \mu \h \nu}^{(2)}(\h x)$ 
gives one antisymmetric tensor $\t B_{\t \mu \t \nu}^{(2)}(\t x) 
\otimes \bf I$ and 22 scalars $\t b_2 (\t x) \otimes \b f_{\b m \b n}(\b y)$, 
whereas, the 
four-form anti-self-dual tensor $\h A_{\h \mu \h \nu \h \rho \h \sigma}^-
(\h x)$ gives 19 self-dual two-forms $\t A_{\t \mu\t \nu}^-(\t x) \otimes
\b f^-_{\b m \b n}(\b y)$ and 3 anti-self-dual two forms 
$\t A_{\t \mu\t \nu}^-(\t x) 
\otimes \b f^+_{\b m \b n}(\b y)$ in six dimensions. Finally, 
the anti-self-dual four-form
will give one scalar $\t a_2(\t x)$ half of which comes from $\t A^-_{\b m
\b n \b p \b q}$
and another half by taking the Hodge-dual of $\t A^-_{\t \mu \t \nu \t \rho
\t \sigma}$. We will not get any gauge field (since $b_3 = 0$) or any 
three-form tensor (since $b_1 = 0$) from it.

Now, we consider the fermionic sector. Here we will make use of the index 
theory on K3 surface. It is well-known that the Dirac operator on K3 surface 
has two zero modes corresponding to two (0, p) forms (0, 0) and (0, 2), 
whereas
the Rarita-Schwinger operator has 40 zero modes corresponding to twice the
number of 20 harmonic (1, 1) forms [23] i.e. $I_{1/2}(K3) = 2$ 
and $I_{3/2}(K3)
= -40$. Here $I$ counts the number of positive chirality minus the number
of negative chirality zero modes. So, the ten dimensional gravitino of negative
chirality $\h \psi_{\h \mu}^{(1)\,-}(\h x)$ will yield one six dimensional 
gravitino of negative chirality $\t \psi_{\t \mu}^{(1)\,-}(\t x) \otimes
\b \eta^+_{1/2}(\b y)$, counted in units of Weyl spinor and 20 positive 
chirality spin 1/2 Weyl fermions $\t \chi^{(1)\,+}(\t x) \otimes
\b \eta^-_{\b m}(\b y)$. Finally, one positive chirality spin 1/2 M-W fermion
$\h \lambda^{(1)\,+}(\h x)$ will give a positive chirality Weyl fermion
$\t \lambda^{(1)\,+}(\t x) \otimes \b \eta^+_{1/2}(\b y)$ in six dimensions.
Thus we have obtained the six dimensional fermions from the NS-R sector. From
the R-NS sector we get the identical spectrum i.e. one gravitino of negative 
chirality and 21 positive chirality spin 1/2 Weyl fermions.

So, the complete spectrum of type IIB theory on K3 consists of
one graviton $\big(\t g_{\t \mu \t \nu}(\t x)\big)$, two gravitinos of negative
chirality $\big(\t \psi_{\t \mu}^{(1)\,-}(\t x), \t \psi_{\t \mu}^{(2)\,-}
(\t x)\big)$,
105 scalars $\big(58\,\t a_1(\t x) + 22\, \t b_1(\t x) + 22\, \t b_2(\t x) +
1\,\t \phi^{(1)}(\t x) + 1\, \t a_2(\t x) + 1\, \t \phi^{(2)}(\t x)\big)$, 21 
self-dual antisymmetric tensors $\big(1\, \t B_{\t \mu \t \nu}
^{(1)\,+}(\t x) +
1\, \t B_{\t \mu \t \nu}^{(2)\,+}(\t x) + 19\,\t A_{\t \mu \t \nu}^+(\t x)
\big)$,
5 anti-self-dual antisymmetric tensors $\big(1\, \t B_{\t \mu \t \nu}
^{(1)\,-}(\t x) +
1\, \t B_{\t \mu \t \nu}^{(2)\,-}(\t x) + 3\,\t A_{\t \mu \t \nu}^-(\t x)
\big)$ and
42 positive chirality spin 1/2 Weyl fermions $\big(20\, \t 
\chi^{(1)\,+}(\t x) +
1\, \t \lambda^{(1)\,+}(\t x) + 20\,\t \chi^{(2)\,+}(\t x) + 1\, \t \lambda
^{(2)\, +}(\t x)\big)$. It can be easily checked that the spectrum is free of 
gravitational anomaly [28] since they satisfy $I_{3/2} : I_{1/2} : I_A = 
-2 : 42 : 16 = 1 : -21 : -8$. Here $I_{3/2}$, $I_{1/2}$ and $I_A$ are 
respectively the gravitational anomalies for spin 3/2 (positive chirality
minus negative chirality) spin 1/2 and antisymmetric tensor (self-dual minus
anti-self-dual) fields. Note that the spectrum has chiral N=2 supersymmetry and
it contains one gravity multiplet $(\t g_{\t \mu \t \nu}, \t \psi^{I\,\alpha}
_{\t \mu}, 5\,\t A_{\t \mu\t \nu}^{(IJ)\,-})$ and 21 antisymmetric tensor 
multiplets $(\t A_{\t \mu\t \nu}^+, \t \lambda^I_{\alpha}, 5\, \t 
\phi^{(IJ)})$
where $I$, $J$ are USp(4) indices and $\alpha$ is the spinor index for which
up(down) means antichiral(chiral) spinors. Also note that the 105 scalars 
parametrize the moduli space O(21, 5)/(O(21) $\times$ O(5)). Identical
spectrum has also been obtained from M-theory compactification on the orbifold
T$^5$/Z$_2$ [18, 19].

We now follow this procedure to obtain the massless spectrum of the two 
dimensional reduction of type IIB theory on K3 $\times$ K3. We note that in
2d, vectors or higher rank tensors do not have any propagating degree of 
freedom and therefore, we will not count them in the spectrum. Also, both
the graviton and the gravitino have formally $-1$ degree of freedom and so,
graviton has to be compensated by a scalar wheras the gravitinos would have
to be compensated by spin 1/2 M-W fermions. Keeping these in mind, we find 
that the six dimensional graviton will give a 2d graviton $\t g_{\t \mu\t \nu}
(\t x)\rightarrow g_{\mu\nu}(x) \otimes {\bf I}$ 
and 58 scalars $\t g_{\t \mu\t \nu}
(\t x)\rightarrow c_1(x) \otimes \b h_{\b m\b n}(\b y)$. 
58 six dimensional scalars
$\t a_1(\t x)$ will give 58 two dimensional scalars $c_2(x) \otimes {\bf I}$.
The dilaton $\t \phi^{(1)}(\t x)$ will yield a single scalar in 2d $\phi_1(x)
\otimes {\bf I}$. The antisymmetric tensor $\t B_{\t \mu\t \nu}^{(1)}(\t x)$
gives 22 two dimensional scalars $c_3(x) \otimes \b f_{\b m \b n}(\b y)$ and
22 six dimensional scalars $\t b_1(\t x)$ give another set of 22 scalars
$c_4(x) \otimes {\bf I}$ in 2d. Similarly, in the R-R sector, we get from
$\t \phi^{(2)}(\t x)$, one two dimensional scalar $\phi_2(x) \otimes {\bf I}$
and from $\t B_{\t \mu\t \nu}^{(2)}(\t x)$, we get 22 scalars $c_5(x) 
\otimes \b f_{\b m \b n}(\b y)$ and 22 six dimensional scalars $\t b_2(\t x)$
yield one more set of 22 scalars $c_6(x) \otimes {\bf I}$. Now from 19 
self-dual antisymmetric two-forms $\t A_{\t \mu \t \nu}^+(\t x)$ we get 
19 $\times$ 19 = 361 chiral bosons $A_1^-(x) \otimes \b f_{\b m \b n}^-(\b y)$
and 19 $\times$ 3 = 57 antichiral bosons $A_1^+(x) \otimes \b f_{\b m \b n}
^+(\b y)$ in 2d. Similarly, from 3 six dimensional anti-self-dual antisymmetric
two-forms $\t A_{\t \mu \t \nu}^-(\t x)$ we obtain 3 $\times $ 19 = 57
antichiral bosons $A_2^+(x) \otimes \b f_{\b m \b n}^-(\b y)$ and 3 $\times$ 3
= 9 chiral bosons $A_2^-(x) \otimes \b f_{\b m \b n}^+(\b y)$. Finally, from
the six dimensional scalar $\t a_2(\t x)$, we get one more scalar $c_7(x)
\otimes {\bf I}$. So, collecting all the massless particles in the bosonic 
sector we have one graviton $g_{\mu\nu}(x)$, 577 chiral bosons
$\big(58\, c_1^-(x) + 58\, c_2^-(x) + 1\, \phi_1^-(x) + 22\,c_3^-(x)+ 22\, 
c_4^-(x) + 1\, 
\phi_2^-(x) + 22\, c_5^-(x) + 22\, c_6^-(x) + 361\, A_1^-(x) + 9\, A_2^-(x)
+ 1\, c_7^-(x)\big)$ and 321 antichiral bosons $\big(58\, c_1^+(x) 
+ 58\, c_2^+(x) 
+ 1\, \phi_1^+(x) + 22\, c_3^+(x) + 22\, c_4^+(x) + 1\,
\phi_2^+(x) + 22\, c_5^+(x) + 22\, c_6^+(x) + 57\, A_1^+(x) + 57\, A_2^+(x)
+ 1\, c_7^+(x)\big)$. In counting the bosons, note that, we have split the 2d
bosons into chiral and antichiral components. Also note that if we compensate
$-1$ degree of freedom of the two dimensional graviton by a scalar, we will
be left with 576 chiral and 320 antichiral bosons.

We next turn our attention to the fermionic sector. In this case we will apply
exactly the same procedure as the six dimensional reduction on K3. Since in 2d,
we have M-W spinors we will count the spinors in terms of them unlike the case 
in 6d, where they were counted in terms of Weyl spinors. Since the Dirac 
operator and the Rarita-Schwinger operator on K3 has 2 and 40 zero modes 
respectively, we get from two six dimensional gravitinos $\t \psi_{\t \mu}
^{(1)\,-}(\t x)$ and $\t \psi_{\t \mu}^{(2)\,-}(\t x)$, eight two dimensional
gravitinos of negative chirality $\psi_\mu^{(1)\,-}(x) \otimes \b \eta_{1/2}^+
(\b y)$ and $\psi_\mu^{(2)\,-}(x) \otimes \b \eta_{1/2}^+(\b y)$. Note that
there is a pair of $\b \eta_{1/2}^+(\b y)$ and each Weyl spinor splits up into
two M-W spinors. Also, we get 160 spin 1/2 M-W fermions of positive chirality
$\chi^{(3)\,+}(x) \otimes \b \eta_{\b m}^-(\b y)$ and 
$\chi^{(4)\,+}(x) \otimes 
\b \eta_{\b m}^-(\b y)$. Note here, that there are 40 $\b \eta_{\b m}^-(\b y)$
and Weyl spinors split up into two M-W spinors. Also, from 40 positive 
chirality spin 1/2 Weyl fermions $\t \chi^{(1)\,+}(\t x)$ and 
$\t \chi^{(2)\,+}(\t x)$, we get 160 spin 1/2 M-W fermions $\chi^{(1)\,+}(x)
\otimes \b \eta_{1/2}^+(\b y)$ and $\chi^{(2)\,+}(x) \otimes \b \eta_{1/2}^+
(\b y)$ of positive chirality in 2d. Finally, from two Weyl fermions of 
positive chirality $\t \lambda^{(1)\,+}(\t x)$ and $\t \lambda^{(2)\,+}(\t x)$, 
we get eight positive chirality spin 1/2 M-W fermions $\lambda^{(1)\,+}(x)
\otimes \b \eta_{1/2}^+(\b y)$ and $\lambda^{(2)\,+}(x) 
\otimes \b \eta_{1/2}^+
(\b y)$. So, now counting all the fermions including NS-R and R-NS sectors
we have eight gravitinos of negative chirality $\big(4\,\psi_\mu^{(1)\,-}(x) +
4\,\psi_\mu^{(2)\,-}(x)\big)$ and 328 positive chirality spin 1/2 M-W fermions
$\big(80\, \chi^{(3)\,+}(x) + 80\, \chi^{(4)\,+}(x) + 80\, \chi^{(1)\,+}(x)
+ 80\, \chi^{(2)\,+}(x) + 4\, \lambda^{(1)\,+}(x) + 4\, \lambda^{(2)\,+}(x)
\big)$.
Note that we have a chiral N=8 supersymmetry since we got eight gravitinos
in 2d, which is consistent, because we started out with a chiral N=2 
supersymmetric theory in ten dimensions which would have given a chiral N=32
supersymmetry if we considered a toroidal compactification. But since K3 is 
half-flat it preserves only half of the original supersymmetry and therefore 
on K3 $\times $ K3 reduction only 1/4 th of the original supersymmetry should 
remain and thus we got a chiral N=8 supersymmetry in 2d. In terms of chiral
N=8 supermultiplets the spectrum can be arranged as a gravity multiplet
$(g_{\mu\nu},\, 8\,\psi_\mu^-,\, 8\, \lambda^+, \phi)$ and 40 matter multiplets
$(8\,\lambda^+,\, 8\, \phi^+)$. The rest of the chiral bosons $576\,\phi^-$
remain inert under supersymmetry. We have generically denoted the scalars as
$\phi$ and spin 1/2 M-W fermions as $\lambda$.

Now we show that the 2d model we have obtained is free of gravitational 
anomaly [28]. The gravitational anomaly associated with spin 3/2 field (chiral
minus antichiral) is given by $I_{3/2} = \frac{23}{24} p_1$, that of spin 1/2 
field is $I_{1/2} = -\frac{1}{24}p_1$ and for the chiral minus antichiral 
boson is $I_s = -\frac{1}{24}p_1$, with $p_1$ being the anomaly polynomial. 
One can form an anomaly-free combination from them as $I_{3/2} - m I_{1/2}
+ (23+m) I_s = 0$, where $m$ is any integer. So, the spectrum will be 
anomaly-free if they satisfy $I_{3/2} : I_{1/2} : I_s = 1 : -m : (23+m)$.
Note here that the fermions in this formula are complex (Weyl) fermions and
their number will be half of the M-W fermions we have counted. The two 
dimensional spectrum we have obtained satisfies $I_{3/2} : I_{1/2} : I_s
= -4 : 164 : -256 = 1 : -41 : (23+41)$. We thus have a consistent two 
dimensional compactifications of type IIB theory on K3 $\times$ K3.

\vspace{1cm}

\begin{large}
\noindent{\bf III. An Orbifold Projection:}
\end{large}

\vspace{.5cm}

In this section, we consider an orbifold projection of the above two 
dimensional model of type IIB theory on K3 $\times$ K3. The orbifold group
of transformation we consider consists of the product $(-1)^{F_L}\,
\cdot\,\sigma$, where $F_L$ is the space time fermion number operator in the
left moving sector of the world-sheet and $\sigma$ denotes the involution
on the K3 surface. Since type IIB theory is invariant under these operations
we find the orbifold model by projecting out this symmetry where only the
massless states which remain invariant under these operations will be retained.
We will first compute the massless spectrum originating in the untwisted sector
and then point out how the twisted sector states could be obtained in analogy
with ref.[29]. Finally, we will show that the spectrum thus obtained is free
of gravitational anomaly.

By constructing a special K3 surface it has been shown in ref.[30] how the 
involution $\sigma$ acts on various harmonic forms. Making use of the 
Lefschetz fixed point theorem, it is found that out of twenty harmonic
(1, 1) forms only twelve remain invariant and eight change sign under the 
action of $\sigma$. Out of these twelve, eleven are anti-self-dual and the
remaining one Kahler (1, 1) form and the other two (2, 0) and (0, 2) two-forms
are self-dual. Thus we have $b_2^+ = 3$, $b_2^- = 11$, which remain invariant
and the other eight $b_2^-$ change sign. Now we consider the action of
$(-1)^{F_L}$ on various fields of type IIB theory. Since $(-1)^{F_L}$ has the
effect of changing the sign of a fermion in the left moving sector, it is 
clear from (1), that all the states in the R-R sector will change sign, 
leaving the NS-NS sector invariant. Whereas the fermions originating in the 
R-NS sector will change sign leaving the NS-R sector invariant. Thus 
summarizing:
\eqabegin
(-1)^{F_L} : & & \h g_{\h \mu \h \nu} \rightarrow 
\h g_{\h \mu \h \nu}; \quad \h \phi^{(1)} \rightarrow 
\h \phi^{(1)}; \quad  \h B_{\h \mu \h \nu}^{(1)} \rightarrow
\h B_{\h \mu \h \nu}^{(1)}\nn\\
& & \h \phi^{(2)} \rightarrow -\h \phi^{(2)}; \quad \h B_{\h \mu \h \nu}^{(2)} 
\rightarrow -\h B_{\h \mu \h \nu}^{(2)}; \quad \h A_{\h \mu \h \nu \h \rho
\h \sigma}^- \rightarrow - \h A_{\h \mu \h \nu \h \rho \h \sigma}^-\nn\\
& & \h \psi^{(1)\, -}_{\h \mu} \rightarrow \h \psi^{(1)\, -}_{\h \mu};
\quad \h \lambda^{(1)\, +} \rightarrow \h \lambda^{(1)\, +}\nn\\
& & \h \psi^{(2)\, -}_{\h \mu} \rightarrow -\h \psi^{(2)\, -}_{\h \mu};
\quad \h \lambda^{(2)\, +} \rightarrow -\h \lambda^{(2)\, +}
\eqaend 
It is now straightforward to compute the massless spectrum of this orbifold
model which remain invariant under the combined operations $(-1)^{F_L}\,
\cdot\,\sigma$. We have seen before that the ten dimensional metric gives 58
scalars on K3 reduction corresponding to the number of gauge invariant
deformations of the metric. In this case, the number of scalars which would
remain invariant under the involution $\sigma$ would be $b_2^+ \times b_2^- 
+ 1 = 34$. So, on K3 $\times$ K3 reduction we will get (34 + 34) scalars,
denoted by $d_1(x)$ from $\h g_{\h \mu \h \nu}(\h x)$ which will remain 
invariant under $ (-1)^{F_L}\,\cdot\,\sigma$. We will also get a two 
dimensional graviton $g_{\mu\nu}(x)$. From $\h \phi^{(1)}(\h x)$ we will get
one scalar $d_2(x)$ in 2d. Also, from $\h B_{\h \mu \h \nu}^{(1)}(\h x)$,
we get (14 +14) scalars $d_3(x)$, since 14 of the 22 harmonic two-forms
are invariant. $\h \phi^{(2)}(\h x)$ will not give a scalar since the (0, 0) 
form is invariant under $\sigma$. From $\h B_{\h \mu \h \nu}^{(2)}(\h x)$
we get (8+ 8) scalars $d_4(x)$, since eight harmonic (1, 1) forms change sign
under $\sigma$. To count the number of invariant massless states from 
$\h A_{\h \mu \h \nu \h \rho \h \sigma}^-(\h x)$, we first note that there 
will not be any scalar in 6d or in 2d coming from it since it changes sign
under $(-1)^{F_L}$. However, scalars would be formed out of the six dimensional
two-forms. Under first K3 reduction we found that there are 19 self-dual 
two-forms and 3 anti-self-dual two-forms out of which 8 self-dual two-forms
change sign under $\sigma$. So, in the next K3 reduction we will obtain 
8 $\times$ 11 = 88 chiral bosons $d_5^-(x)$ and 8 $\times$ 3 = 24 antichiral
bosons $d_6^+(x)$. Also, from the 11 self-dual two-forms which remained 
invariant under $\sigma$ in the first K3 reduction, we will get 11 $\times$ 8
= 88 chiral bosons $d_7^-(x)$ and from 3 anti-self-dual two-forms, which also 
remained invariant under $\sigma$ in the first K3 reduction, we will get 
3 $\times$ 8 = 24 antichiral bosons $d_8^+(x)$. So, counting all the states 
in the bosonic sector we have apart from a two dimensional graviton, 289
chiral bosons $\big(68\, d_1^-(x) + 1\, d_2^-(x) + 28\, d_3^-(x)+ 16\,
d_4^-(x) + 88\, d_5^-(x) + 88\, d_7^-(x)\big)$ and 161 antichiral bosons 
$\big(68\, d_1^+(x) + 1\, d_2^+(x) + 28\, d_3^+(x)+ 16\,
d_4^+(x) + 24\, d_6^+(x) + 24\, d_8^+(x)\big)$. If we compensate 
$-1$ degree of
freedom of two dimensional graviton by a scalar we will 
be left with 288 chiral and 160 antichiral bosons.

In order to count the fermionic states, we make use of the index
theory of K3 surface as before. Corresponding to two invariant (0, p) 
forms (0, 0) and
(0, 2), the Dirac operator has two zero modes and therefore we get 4 
gravitinos from $\h \psi^{(1)\,-}_{\h \mu}(\h x)$ of negative chirality in 2d.
Also, from the 12 invariant harmonic (1, 1) forms we get (48 + 48) 
spin 1/2 M-W fermions of positive chirality $\chi^{(1)\,+}(x)$ in 2d. From 
$\h \lambda^{(1)\,+}(\h x)$ we get 4 spin 1/2 M-W fermions of positive 
chirality $\lambda^{(1)\,+}(x)$. The gravitino 
from R-NS sector $\h \psi^{(2)\,
-}_{\h \mu}(\h x)$ does not give any gravitino in 2d since it changes sign
under $(-1)^{F_L}$, whereas it gives (32 + 32) spin 1/2 M-W fermions of 
positive chirality $\chi^{(2)\,+}(x)$, corresponding to the 8 harmonic (1, 1)
forms which change sign under $\sigma$. Finally, $\h \lambda^{(2)\,+}(\h x)$
does not give rise to any M-W spin 1/2 fermions. So, altogether we have 4
gravitinos of negative chirality and 164 spin 1/2 M-W fermions of positive 
chirality $\big(96\, \chi^{(1)\,+}(x) + 4\, \lambda^{(1)\,+}(x) + 64\, 
\chi^{(2)\,+}(x)\big)$. Taking into account only the untwisted sector 
states, we 
verify that the spectrum is already anomaly-free since in this case we have,
$I_{3/2} : I_{1/2} : I_s = -2 : 82 : -128 = 1: -41 : 64 = 1 : -41 : (23+41)$.
We also note that this model has a chiral N=4 supersymmetry because of the
presence of 4 gravitinos of negative chirality.   

It is known from M-theory compactifications [21] that the condition of 
gravitational
anomaly cancellation by itself is not always powerful enough to determine the
massless spectrum completely. Even in the case of some orientifold models
of type IIB theory on K3, the untwisted sector itself becomes 
anomaly-free [20].
So, one has to rely on some other principle to obtain the massless states from
the twisted sector. The orbifold model of type IIB theory on K3 $\times$ K3, 
that we considered is also anomaly-free, as we have seen, if we consider only
the untwisted sector states. In order to find the twisted sector states, we 
will follow closely the arguments given by Sen in ref.[21,29]. We first note 
that orbifold of K3 $\times$ K3 has 64 fixed points --- eight from each K3.
Near each of these fixed points the space would look like T$^8$/$(-1)^{F_L}\,
\cdot\, {\cal I}_8$ and therefore the physics in the neighborhood of those
fixed points would be the same for type IIB theory either on (K3 $\times$ K3)
/$(-1)^{F_L}\,\cdot\, \sigma$ or on T$^8$/$(-1)^{F_L}\,\cdot\, {\cal I}_8$. 
Here ${\cal I}_8$ has the effect of changing the sign of all the coordinates
of T$^8$. Fortunately, the twisted sector states of type IIB theory on 
T$^8$/$(-1)^{F_L}\,\cdot\, {\cal I}_8$ have already been computed by Sen in
the second reference [29] just mentioned. It has been argued there by
converting this theory to type IIA theory on T$^8$/${\cal I}_8$ with an
$R\rightarrow 1/R$ transformation on one of the circles of T$^8$ and 
through a
tadpole calculation [31, 29] that the twisted sector 
states of this theory live on the
sixteen elementary type IIB strings (NS one-branes) each of which supports
a vector multiplet of two dimensional N=16 supersymmetry algebra with eight
scalar components labelling the location of these strings on the internal
space. Since the orbifold of K3 $\times$ K3 has 1/4th as many fixed points
as the orbifold of T$^8$, the twisted sector states in this case live on
four elementary type IIB strings. It is, therefore, 
clear that just like what happens for the orientifold model of type IIB 
theory on K3, the spectrum is anomaly-free, even when we include the twisted
sector states, as expected.
\vspace{1cm}
\vfil
\eject
\begin{large}
\noindent{\bf IV. An Orientifold Projection:}
\end{large}

\vspace{.5cm}

We now consider an orientifold model [32--35] of type 
IIB theory on K3 $\times$
K3. The orientifold group that we consider consists of a product of the 
involution on K3 surface and the orientation reversal of the world-sheet
which is denoted as $\Omega$. The orientation reversal basically interchanges
the left moving modes of the string with the right moving modes [36]. Under 
the
operation $\Omega$, it is clear from (1), that the graviton and the dilaton
in the NS-NS sector will remain invariant, whereas the antisymmetric tensor
$\h B_{\h \mu\h \nu}^{(1)}(\h x)$ will change sign since it involves an 
antisymmetric combination of the left moving and the right moving oscillators.
On the other hand, the bosonic fields in the R-R sector which are formed
out of symmetric combination of the left-moving and the right-moving modes 
will change sign because the vertex operator contains a product of spin-fields
in the left and the right-moving sectors and gives a sign under $\Omega$. So, 
the scalar $\h \phi^{(2)}(\h x)$ and $\h A_{\h \mu \h \nu \h \rho \h \sigma}
^-(\h x)$ will change sign whereas $\h B_{\h \mu\h \nu}^{(2)}(\h x)$ will 
remain invariant. In the fermionic sector, the fermions formed out of 
the positive combination of the NS-R and the R-NS sector $\bf(8_v^L \otimes
8_c^R + 8_c^L \otimes 8_v^R)$ remain invariant whereas the fermions formed
out of negative combination $\bf(8_v^L \otimes 8_c^R - 8_c^L \otimes 8_v^R)$
change sign. So, summarizing:
\eqabegin
\Omega : & & \h g_{\h \mu \h \nu} \rightarrow
\h g_{\h \mu \h \nu}; \quad \h \phi^{(1)} \rightarrow
\h \phi^{(1)}; \quad  \h B_{\h \mu \h \nu}^{(1)} \rightarrow
-\h B_{\h \mu \h \nu}^{(1)}\nn\\
& & \h \phi^{(2)} \rightarrow -\h \phi^{(2)}; \quad \h B_{\h \mu \h \nu}^{(2)}
\rightarrow \h B_{\h \mu \h \nu}^{(2)}; \quad \h A_{\h \mu \h \nu \h \rho
\h \sigma}^- \rightarrow - \h A_{\h \mu \h \nu \h \rho \h \sigma}^-\nn\\
& & (\h \psi^{(1)\, -}_{\h \mu} + \h \psi^{(2)\, -}_{\h \mu}) \rightarrow 
(\h \psi^{(1)\, -}_{\h \mu} + \h \psi^{(2)\, -}_{\h \mu});
\quad (\h \psi^{(1)\, -}_{\h \mu} - \h \psi^{(2)\, -}_{\h \mu}) \rightarrow
-(\h \psi^{(1)\, -}_{\h \mu} - \h \psi^{(2)\, -}_{\h \mu})\nn\\
& & (\h \lambda^{(1)\, +} + \h \lambda^{(2)\, +}) \rightarrow
(\h \lambda^{(1)\, +} + \h \lambda^{(2)\, +});
\quad (\h \lambda^{(1)\, +} - \h \lambda^{(2)\, +}) \rightarrow
-(\h \lambda^{(1)\, +} - \h \lambda^{(2)\, +})
\eqaend
As in the orbifold model we will first compute the states in the untwisted
sector under the combined operation $\Omega\,\cdot\,\sigma$ and then mention,
in anlogy with ref.[29], how to compute the twisted sector states. 
We again find
that this orientifold model is free of gravitational anomaly. 

The counting of states in this case proceeds exactly the same way as in
the orbifold model with a few differences. As mentioned before, the ten
dimensional graviton $\h g_{\h \mu\h \nu}(\h x)$ gives a two dimensional
graviton and (34 + 34) scalars $e_1(x)$ from two K3 which remains invariant
under $\sigma$. We get one more scalar $e_2(x)$ from $\h \phi^{(1)}(\h x)$
in 2d. Now since $\h B_{\h \mu\h \nu}^{(1)}(\h x)$ changes sign under $\Omega$
we will get (8 + 8) scalars $e_3(x)$ from the reduction on K3 $\times$ K3,
corresponding to the eight harmonic (1, 1) forms which change sign under
$\sigma$. In the R-R sector, $\h \phi^{(2)}(\h x)$ will not give a scalar, 
whereas from $\h B_{\h \mu\h \nu}^{(2)}(\h x)$, we get (14 + 14) scalars
$e_4(x)$ corresponding to 14 two forms on K3 that remain invariant under
$\sigma$. Finally, the counting of states from $\h A_{\h \mu \h \nu \h \rho 
\h \sigma}^-(\h x)$ is exactly the same as in the orbifold model and therefore,
we get (8 $\times$ 11 + 11 $\times$ 8) = 176 chiral bosons $e_5^-(x)$ and 
(8 $\times$ 3 + 3 $\times$ 8) = 48 antichiral bosons $e_6^+(x)$ in 2d.
Collecting all the bosonic states we have one graviton $g_{\mu\nu}(x)$, 289
chiral bosons $\big(68\, e_1^-(x) + 1\, e_2^-(x) + 16\, e_3^-(x)+ 28\,
e_4^-(x) + 176\, e_5^-(x)\big)$ and 161 antichiral bosons $\big(68\, e_1^+
(x) + 1\, e_2^+(x) + 16\, e_3^+(x)+ 28\, e_4^+(x) + 48\, e_6^+(x)\big)$.

The massless spectrum for the fermionic sector can again be obtained by 
applying index theory on K3 surface. Thus, we get from the combination
$(\h \psi^{(1)\, -}_{\h \mu} + \h \psi^{(2)\, -}_{\h \mu})(\h x)$, four
M-W gravitinos $\psi_\mu^-(x)$ of negative chirality in 2d, corresponding
to the two invariant (0, p) forms on K3. Corresponding to 12 
invariant harmonic
(1, 1) forms, we get (48 + 48) spin 1/2 M-W fermions of positive chirality
$\chi^{(1)\,+}(x)$ in 2d. Similarly, we also get 4 spin 1/2 M-W fermions of
positive chirality from $(\h \lambda^{(1)\, +} + \h \lambda^{(2)\, +}) 
(\h x)$, 
denoted as $\lambda^{(1)\,+}(x)$, corresponding to two invariant (0, p) forms.  
Finally, from $(\h \psi^{(1)\, -}_{\h \mu} - 
\h \psi^{(2)\, -}_{\h \mu})(\h x)$, we get (32 + 32) spin 1/2 M-W fermions of 
positive chirality, denoted as $\chi^{(2)\,+}(x)$, corresponding to 8 harmonic
(1, 1) forms which change sign under $\sigma$. Collecting all the fermionic 
states we have four negative chirality gravitinos $\psi^-_{\mu}(x)$ and 164
spin 1/2 M-W fermions of positive chirality $\big(96\,\chi^{(1)\,+}(x) + 4\,
\lambda^{(1)\,+}(x) + 64\, \chi^{(2)\,+}(x)\big)$. Thus we again have a chiral 
N=4 supersymmetry and the cancellation of
the gravitational anomaly is exactly the same as shown before.

Now in order to construct the twisted sector states, we again use the similar
argument as was considered for the orbifold model. In this case, the space 
near the 64 fixed points would look like T$^8$/$\Omega\,\cdot\,{\cal I}_8$.
The twisted sector states of type IIB theory on T$^8$/$\Omega\,
\cdot\,{\cal I}_8$ have also been obtained by Sen [29]. 
In this case, it has been
found that the twisted sector states live on 16 R-R one-branes 
(type IIB strings) and each of them supports a vector multiplet of
two dimensional N=16 
supersymmetry algebra in which the eight scalars label the positions of the 
R-R one-branes on the internal space. Therefore, viewed in terms of
type IIB theory on (K3 $\times$ K3)/$\Omega\,\cdot\,{\cal I}_8$, the twisted
sector states live on four R-R one-branes and thus, once again we find 
that the gravitational 
anomaly cancels even if we include the twisted sector states as expected. The 
massless spectrum matches precisely for both the orbifold and the orientifold 
models including the untwisted as well as the twisted sector states. Note that
this is quite expected since under the ten dimensional SL(2, Z) symmetry of 
type IIB theory, $(-1)^{F_L}$ gets precisely converted to the 
world-sheet parity
transformation $\Omega$. 

We here briefly point out how the M-theory duals of the various type IIB
compactifications, we have considered in this paper, can be obtained by
using a chain of duality arguments [21]. Let us first consider the type IIB
theory on K3 $\times$ K3. By choosing a special K3 surface, this theory
is equivalent to type IIB theory on (K3 $\times$ T$^4$)/$\{1, {\cal I}_4\}$ 
and then making an R $\rightarrow$ 1/R duality transformation on one of
the circles of T$^4$, we convert this to type IIA theory where ${\cal I}_4$
changes to $(-1)^{F_L}\,\cdot\, {\cal I}_4$. So, the above model is dual
to type IIA theory on (K3 $\times$ T$^4$)/$\{1, (-1)^{F_L}\,\cdot\,
{\cal I}_4\}$. Since M-theory on S$^1$ is the strong coupling limit of type
IIA string theory when the radius of the circle is large, this type IIA model
is dual to M-theory on (K3 $\times$ T$^5$)/$\{1, {\cal J}_5\}$, where
${\cal J}_5$ has the effect of changing the sign of all the coordinates of
T$^5$ and the antisymmetric three-form present in M-theory\footnote[1]{
This duality has also been conjectured in [19].}. It is easy to
verify that ${\cal J}_1$ in M-theory has the same effect as $(-1)^{F_L}$
in type IIA theory. The bosonic sector of M-theory compactification on
(K3 $\times$ T$^5$)/$\{1, {\cal J}_5\}$ has been obtained in ref.[37] and
indeed was found to have the same spectrum as type IIB on K3 $\times$ K3.
Next we consider the orbifold model of type IIB theory on 
(K3 $\times$ K3)/$\{1, (-1)^{F_L}\,\cdot\,\sigma\}$. Since the orientifold 
model we have considered in this section is dual to the orbifold model, they
will give the identical M-theory dual. By using a chain of duality arguments
it has been shown in ref.[21] that the type IIB theory on K3/$\{1, (-1)^{F_L}
\,\cdot\,\sigma\}$ is dual to type IIA theory on the same orbifold where
the massless states from the twisted sector get interchanged with those of
the untwisted sector. Since $(-1)^{F_L}$ in type IIA theory has the same effect
as ${\cal J}_1$ in M-theory, we conclude that type IIB theory on 
(K3 $\times$ K3)/$\{1, (-1)^{F_L}\,\cdot\,\sigma\}$ is dual to M-theory 
compactification on (K3 $\times$ K3 $\times$ S$^1$)/$\{1, {\cal J}_1\,\cdot\,
\sigma\}$. It will be interesting to verify how the massless spectrum precisely
arises in this M-theory compactification. Certain relevant two dimensional
compactifications of M-theory have been considered in [38].
 
\vspace{1 cm}

\begin{large}
\noindent{\bf V. Conclusions:}
\end{large}

\vspace{.5cm}

To summarize, we have considered in this paper a two dimensional reduction of
type IIB theory on K3 $\times$ K3. We found that this gives a consistent
compactification of type IIB theory since the resulting spectrum is free of
gravitational anomaly. We have also considered an orbifold and an orientifold
projection of the above model and computed the massless spectrum for both the
untwisted and the twisted sectors. We found that the orbifold and the 
orientifold models have chiral N=4 supersymmetry and the resulting spectrum 
in the two cases match precisely including the untwisted and the twisted sector
states. We have also shown that in both cases, the models are free of 
gravitational anomaly. The matching of the spectrum for these models is a 
consequence of the SL(2, Z) symmetry of the ten dimensional type IIB theory.
We have also briefly pointed out the M-theory duals of the various type IIB
compactifications considered in this paper. By considering other 
orbifold/orientifold of type IIB theory on K3 $\times$ K3, it will be 
interesting to see how to obtain models with less number of supersymmetries
and study their M-theory as well as F-theory duals.
\vspace{1cm}

\begin{large}

\noindent{\bf Acknowledgements:}

\end{large}

\vspace{.5cm}

I would like to thank S. Panda for discussions at an early stage of this 
work. This work is supported in part by the Spanish Ministry of Education
(MEC) fellowship.

\vspace{1cm}

\begin{large}

\noindent{\bf References:}
\end{large}

\vspace{.5cm}
\baselineskip 12pt
\begin{enumerate}
\item M. Duff, R. R. Khuri and J. X. Lu, Phys. Rep. 259 (1995) 213.
\item A. Sen, Int. J. Mod. Phys. A9 (1994) 3707.
\item E. Witten, \np 443 (1995) 85. 
\item C. Hull and P. Townsend, \np 438 (1995) 109. 
\item J. H. Schwarz, \pl 367 (1996) 97.
\item P. Horava and E. Witten, \np 460 (1996) 506.
\item F. Aldabe, {\it Heterotic and Type I Strings from Twisted 
Supermembranes}, hep-th/9603183.
\item O. Aharony, {\it String Theory Dualities from M-theory}, hep-th/9604103.
\item M. J. Duff, R. Minasian and E. Witten, {\it Evidence for Heterotic/
Heterotic Duality}, hep-th/9601036.  
\item B. S. Acharya, {\it N=1 Heterotic/M-Theory Duality and Joyce Manifolds},
hep-th/9603033.
\item A. Das and S. Roy, {\it On M-Theory and the Symmetries of 
Type II String Effective Actions}, hep-th/9605073.
\item J. H. Schwarz, \pl 360 (1995) 13. 
\item C. M. Hull, \pl 357 (1995) 545. 
\item C. Vafa, {\it Evidence for F-Theory}, hep-th/9602022.
\item S. Sethi, C. Vafa and E. Witten, {\it Constraints on Low Dimensional
String Compactifications}, hep-th/9606122.
\item E. Witten, Int. J. Mod. Phys. A10 (1995) 1247. 
\item P. K. Townsend, \pl 139 (1984) 283.
\item K. Dasgupta and S. Mukhi, {\it Orbifolds of M-Theory}, hep-th/9512196. 
\item E. Witten, {\it Five-Branes and M-theory on an Orbifold}, hep-th/9512219.
\item A. Dabholkar and J. Park, {\it An Orientifold of Type IIB Theory on K3},
hep-th/9602030; {\it Strings on Orientifolds}, hep-th/9604178.
\item A. Sen, {\it M-Theory on (K3 $\times$ S$^1$)/Z$_2$}, hep-th/9602010;
{\it Orbifolds of M-Theory and String Theory}, hep-th/9603113.
\item A. Kumar and K. Ray, {\it M-Theory on Orientifolds of K3 $\times$ S$^1$},
hep-th/9602144.
\item M. A. Walton, Phys. Rev. D37 (1988) 377.
\item M. B. Green, J. H. Schwarz and E. Witten, {\it Superstring Theory},
Vol. I, Cambridge University Press, 1987.
\item T. Eguchi, P. Gilkey and A. Hanson, Phys. Rep. 66 (1980) 213.
\item D. Page, \pl 80 (1978) 55.
\item S. Hawking and C. Pope, \np 146 (1978) 381.
\item L. Alvarez-Gaume and E. Witten, \np 234 (1984) 269.
\item A. Sen, {\it Duality and Orbifolds}, hep-th/9604070.
\item J. Schwarz and A. Sen, \pl 357 (1995) 323.
\item C. Vafa and E. Witten, \np 447 (1995) 261.
\item A. Sagnotti, in Cargese '87, {\it Non-perturbative Quantum Field Theory},
eds. G. Mack et.al. (Pergamon Press, 1988), 521; G. Pradisi and A. Sagnotti,
\pl 216 (1989) 59; M. Bianchi and A. Sagnotti, \pl 247 (1990) 517; \np
361 (1991) 519.
\item P. Horava, \np 327 (1989) 461.
\item J. Dai, R. G. Leigh and J. Polchinski, Mod. Phys. Lett A4 (1989) 2073.
\item E. Gimon and J. Polchinski, {\it Consistency Conditions for Orientifolds 
and D-Manifolds}, hep-th/9601038.
\item J. Polchinski, S. Chaudhuri and C. V. Johnson, {\it Notes on D-Branes},
hep-th/9602052.
\item B. S. Acharya, {\it M-theory Compactification and Two-Brane/Five-Brane
Duality}, hep-th/9605047.
\item A. Kumar and K. Ray, {\it Compactifications of M-Theory to Two
Dimensions}, hep-th/9604164.
\end{enumerate}

\vfil

\vfil
\eject 

\end{document}